\documentclass[aps,prd,letterpaper,twocolumn,preprintnumbers,nofootinbib]{revtex4-2}

\usepackage{amsmath,amssymb}
\usepackage{graphicx}
\usepackage{footmisc}
\usepackage{subfig}
\usepackage[hyperfootnotes=false,colorlinks,citecolor=blue]{hyperref}

\def \L {\mathcal{L}} 

\def \vec#1{{\boldsymbol{#1}}}
\newcommand{\dd}{{\rm d}}
\newcommand{\ii}{{\rm i}}

\allowdisplaybreaks

\begin{document}

\title{Revisiting the Friedberg--Lee--Sirlin soliton model}

\author{Julian Heeck}
\email[Email: ]{heeck@virginia.edu}
\thanks{ORCID: \href{https://orcid.org/0000-0003-2653-5962}{0000-0003-2653-5962}.}
\affiliation{Department of Physics, University of Virginia,
Charlottesville, Virginia 22904-4714, USA}

\author{Mikheil Sokhashvili}
\email[Email: ]{ms2guc@virginia.edu}
\thanks{ORCID: \href{https://orcid.org/0000-0003-0844-7563}{0000-0003-0844-7563}.}
\affiliation{Department of Physics, University of Virginia,
Charlottesville, Virginia 22904-4714, USA}

\hypersetup{
pdftitle={Revisiting the Friedberg-Lee-Sirlin soliton model},   
pdfauthor={Julian Heeck, Mikheil Sokhashvili}
}


\begin{abstract}
Non-topological solitons are localized classical field configurations stabilized by a Noether charge. Friedberg, Lee, and Sirlin proposed a simple renormalizable soliton model in their seminal 1976 paper, consisting of a complex scalar field that carries the Noether charge and a real-scalar mediator. We revisit this model, point out commonalities and differences with $Q$-ball solitons, and provide analytic approximations to the underlying differential equations.
\end{abstract}

\maketitle


\section{Introduction}

Non-topological solitons are localized field configurations that are stabilized by a conserved Noether charge~\cite{Lee:1991ax}. They are solutions to the classical field equations but can be interpreted as bound states within quantum field theory.
Arguably the simplest class of such solitons arises in scalar field theories involving at least two real or one complex scalars, invariant under orthogonal or unitary symmetry groups~\cite{Wick:1954eu,Rosen:1968mfz,Rosen:1969ay,Cutkosky:1954ru,Friedberg:1976me}.

Scalar solitons in which the contributing scalar fields take on approximately constant values inside and vanish outside have been analyzed by Coleman and dubbed $Q$-balls~\cite{Coleman:1985ki}, $Q$ being the Noether charge of the soliton. These solitons allow for simple analytical approximations that have been widely employed and generalized~\cite{PaccettiCorreia:2001wtt,Heeck:2020bau,Lennon:2021zzx,Heeck:2022iky}.
In the large $Q$ -- or thin-wall -- limit, the $Q$-ball energy $E$ is approximately proportional to the charge $Q$ or the $Q$-ball volume $V$, with a prefactor that depends on the underlying scalar potential~\cite{Heeck:2022iky}.\footnote{When the leading prefactor vanishes, one finds instead $E\propto Q^{4/5} \propto V^{2/3}$~\cite{Spector:1987ag,PaccettiCorreia:2001wtt,Heeck:2020bau} in three dimensions.}

As popular as Coleman's $Q$-balls are as examples of non-topological solitons, they are by no means the only possible stable configurations. In fact, one of the earliest and best-known models for a scalar soliton, that of Friedberg, Lee, and Sirlin (FLS)~\cite{Friedberg:1976me}, does not have a thin-wall limit in the sense of Coleman and should therefore not be called a $Q$-ball, despite the qualitatively similar internal structure.
Here, we will revisit the FLS model and provide analytic approximations and numerical cross-checks that give an exhaustive overview of the solitons and their differences with $Q$-balls.

The rest of this paper is organized as follows: in Sec.~\ref{sec:model} we introduce our notation of the FLS model and provide expressions of important macroscopic quantities such as energy and charge. In Sec.~\ref{sec:Thin-wall} we derive approximate analytical solutions for the fields in a thin-wall-like regime, introduce the radii, and calculate integrals that are required for the macroscopic soliton properties. This section is split into two subsections covering heavy and light mediator limits separately. Sec.~\ref{sec:Thick-wall} presents calculations for the thick-wall regime, restricted to the light mediator limit that leads to stable solitons. In Sec.~\ref{sec:stability} we study the soliton stability. We discuss our results in Sec.~\ref{sec:discussion} and finally conclude in Sec.~\ref{sec:conclusion}.

\section{Friedberg--Lee--Sirlin Model}
\label{sec:model}

In this article, we are going to use the mostly-minus Minkowski metric. The FLS model consists of one complex scalar field $\phi$ and one real mediator field $\chi$, with a Lagrangian~\cite{Friedberg:1976me}
\begin{align}
    \L = |\partial_\mu \phi|^2 + \frac{1}{2}\partial_\mu\chi\partial^\mu\chi - U(|\phi|,\chi)
    \label{eq:GeneralLagrangian}
\end{align}
that is invariant under the $U(1)$ symmetry $\phi\to e^{\ii \alpha} \phi$, $\chi\to\chi$ as well as a $\mathbb{Z}_2$ symmetry $\chi\to -\chi$, $\phi\to\phi$:
\begin{align}
\begin{split}
    U(|\phi|,\chi) = d^2 \chi^2 |\phi|^2 + \frac{1}{8} g^2 \left( \chi^2 - \chi_\text{vac}^2 \right)^2.
    \label{eq:PotentialUFLS}
\end{split}
\end{align}
According to Noether's theorem there is a conserved charge $Q$ due to global $U(1)$ symmetry, which is normalized to $Q(\phi)=1$. Therefore,  $Q$ counts the number of $\phi$ particles. 
The FLS potential is chosen so that the $\mathbb{Z}_2$ symmetry is spontaneously broken. We expand around the true minimum by shifting  $\chi \rightarrow \chi_\text{vac} - \chi$. Next, we switch from $\chi_\text{vac}$ and $g$ to the new parameters $m_\chi$ and $m_\phi$, the particle masses in the true vacuum outside of the soliton. Expanded around the true minimum, the potential takes the form
\begin{align}
    U(|\phi|,\chi) &= m_\phi^2|\phi|^2+\frac{1}{2}m_\chi^2\chi^2-2 d m_\phi \chi |\phi|^2 + d^2 \chi^2 |\phi|^2\nonumber\\ 
    &\quad -\frac{d m_\chi^2}{2m_\phi}\chi^3+\frac{d^2 m_\chi^2}{8 m_\phi^2}\chi^4 \,.
    \label{eq:PotentialU}
\end{align}
$\phi$ does not have any local self-interactions, but the mediator $\chi$  provides attractive interactions that allow for a bound state solution with charge $Q$. The solution with the lowest energy among those with the same $Q$ will be considered stable. The lowest-energy configuration will be spherically symmetrical~\cite{Friedberg:1976me} and have a simple time dependence to evade Derrick's theorem~\cite{Derrick:1964ww}. Hence we introduce the following parametrization:
\begin{align}
    \phi(\Vec{x},t)=\phi_0 e^{\ii \omega t}f(|\Vec{x}|)\,, \quad     \chi(\Vec{x},t)=\chi_0 h(|\Vec{x}|)\,.
\end{align}
Equations \eqref{eq:GeneralLagrangian} and \eqref{eq:PotentialU} lead to equations of motion that depend on the parameters $m_\phi$, $m_\chi$, $d$, $\omega$, $\phi_0$, and $\chi_0$. 

\begin{figure}
\centering
\subfloat[$\kappa = 0.05$, $\eta = 10^{-6}$]{
  \includegraphics[width=42mm]{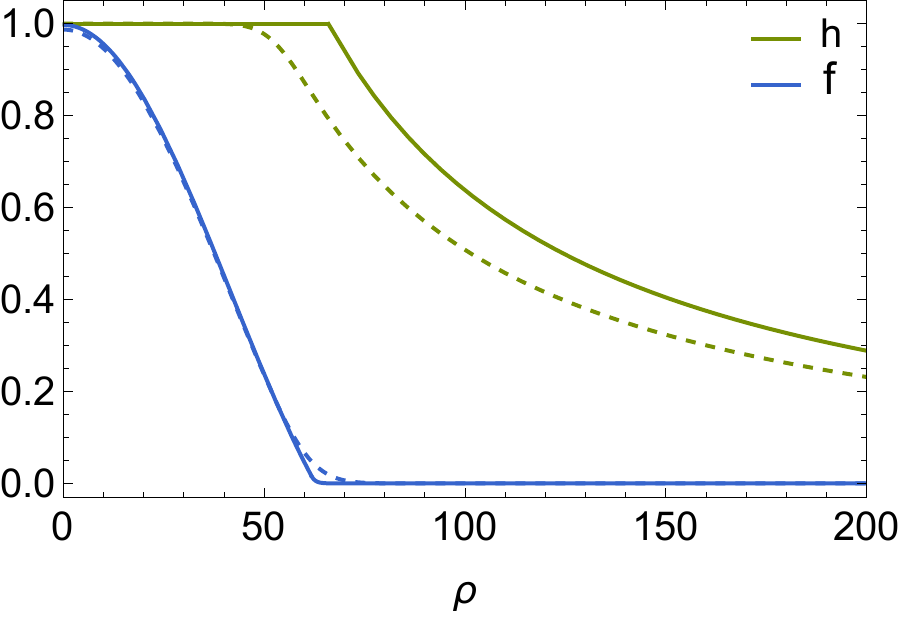}
}
\subfloat[$\kappa = 0.05$, $\eta = 10^{4}$]{
  \includegraphics[width=42mm]{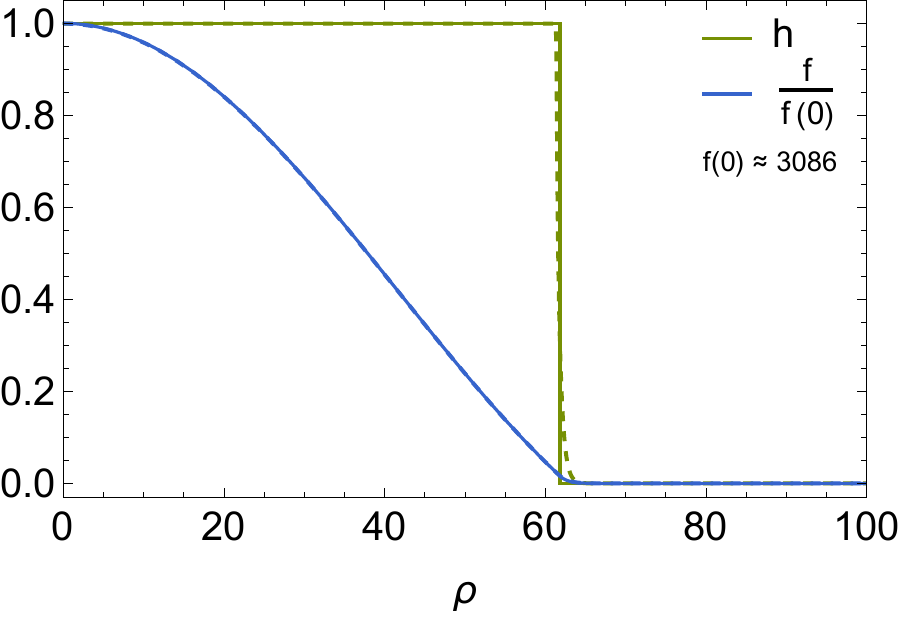}
}
\hspace{0mm}
\subfloat[$\kappa = 0.95$, $\eta = 10^{-6}$]{
  \includegraphics[width=42mm]{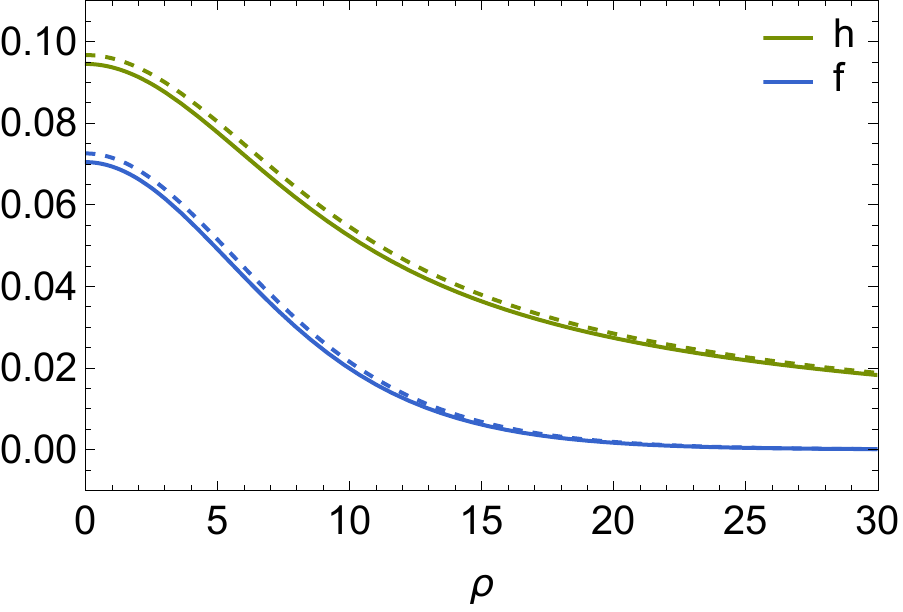}
}
\subfloat[$\kappa = 0.95$, $\eta = 10^{4}$]{
  \includegraphics[width=42mm]{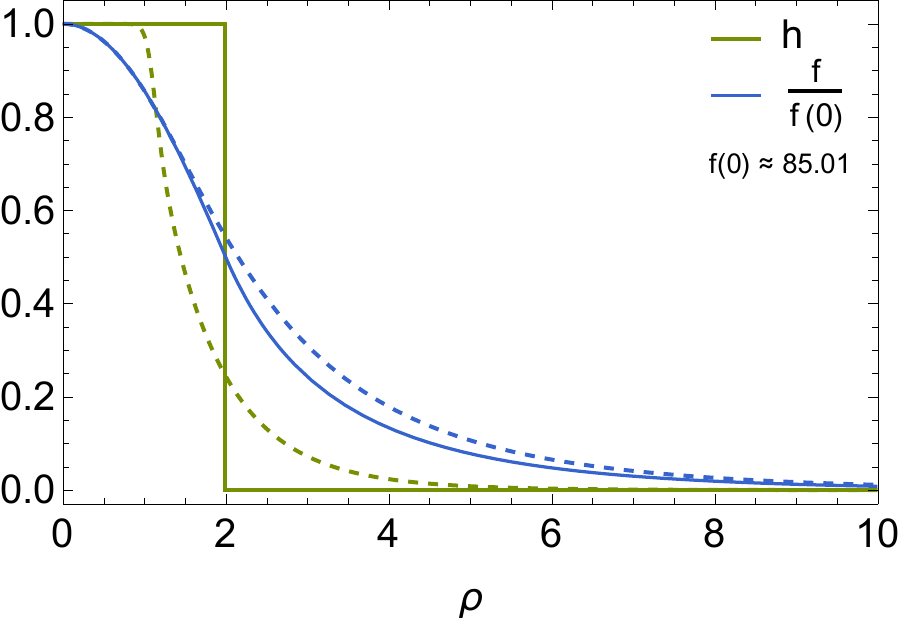}
}
\caption{  $f$ and $h$ profiles for various parameter choices.  
Dashed lines are  numerical results and solid lines represent our theoretical predictions from equations  \eqref{eq:FProfile}, \eqref{eq:F0Heavy}, \eqref{eq:F0Light}, \eqref{eq:ProfilesThick}, and \eqref{eq:HProfileLight}. As we do not have a theoretical prediction for a thick-wall-heavy-$\chi$ case, we used our thin-wall-heavy-$\chi$ formula for case (d).}
\label{fig:PlotProfiles}
\end{figure}

We can simplify the equations by making a coordinate transformation $\vec{x}\to \vec{x}/\lambda$ with a length scale $\lambda$ and furthermore fix the ratio of prefactors via $\phi_0=\chi_0/\sqrt{2}$. This allows us to factor out both $\chi_0$ and $\lambda$. The rescaled one-dimensional Lagrangian is 
\begin{align}
    L = 4 \pi \lambda \chi_0^2 \int \dd\rho \, \rho^2\left[ -\frac{f'^2}{2} - \frac{h'^2}{2} + V(f,h) \right],
\end{align}
where  $\rho\equiv |\vec{x}|/\lambda$ is the dimensionless radial coordinate and the rescaled potential $V$ is
\begin{align}
\begin{split}
    V &= -\frac{\lambda^2}{\chi_0^2}(U-\omega^2|\phi|^2)\\
    &= (\kappa^2-1)\frac{f^2}{2}+f^2h-\frac{f^2h^2}{2}-\frac{h^2}{2} \eta +\frac{h^3}{2} \eta-\frac{h^4}{8} \eta\,.
    \label{eq:PotentialV}
\end{split}
\end{align}
The last equation of eq.~\eqref{eq:PotentialV} is obtained by introducing the dimensionless parameters
\begin{align}
\kappa \equiv \omega/m_\phi\,, \text{  and  } \eta\equiv m_\chi^2/m_\phi^2
\end{align}
and fixing $\lambda\equiv 1/m_\phi$ and $\chi_0\equiv m_\phi/d$. Our reparametrization differs from FLS~\cite{Friedberg:1976me} but leads to the same number of remaining parameters.
The equations of motion 
\begin{align}
\begin{split}
    &f''(\rho)+\frac{2}{\rho}f'(\rho)+\frac{\partial V}{\partial f}=0\,,\\
    &h''(\rho)+\frac{2}{\rho}h'(\rho)+\frac{\partial V}{\partial h}=0
        \label{eq:EOM}
\end{split}
\end{align}
now only depend on $\kappa$ and  $\eta$:
\begin{align}
    &f''+\frac{2}{\rho}f'-f h^2+2 f h+f \kappa ^2-f=0\,, \label{eq:EOMExplicit_f}\\
    &h''+\frac{2}{\rho}h'-f^2 h+f^2 - \eta h + \frac{3\eta}{2}h^2 - \frac{\eta}{2}h^3 =0\,.
    \label{eq:EOMExplicit_h}
\end{align}
As we are looking for localized solutions, the boundary conditions are given by $f'(0)=h'(0)=0$ and $f(\rho\to\infty) = h(\rho\to\infty) = 0$.\footnote{It is worth mentioning that for each solution that ends at $f=0\,\&\,h=0$ there is a corresponding solution that ends at $f=0\,\&\,h=2$. This is easy to verify by observing a symmetry of our potential $V[f,h]$ with respect to the $h=1$ plane. Since the corresponding solutions will be identical to one another by simply replacing $h(\rho)$ with $h_\text{corresponding}(\rho)=2-h(\rho)$, we can restrict ourselves to $h\geq0$ solutions. }
These differential equations are identical in form to those describing false vacuum decay~\cite{Coleman:1977py}, which have been under much more scrutiny than their soliton analogues. However, only few numerical codes can handle the equations at hand, which correspond to two-field vacuum decay into an unbounded regime. 
We have solved the equations numerically both via the shooting method in \texttt{Mathematica} and using the finite-difference method on a compactified grid outlined in Ref.~\cite{Heeck:2020bau}. For $\mathcal{O}(1)$ $\kappa$ and $\eta$ the public code of Ref.~\cite{Sato:2019wpo} has been used for comparison as well.
We show numerical solutions of the two field profiles in Fig.~\ref{fig:PlotProfiles} together with various analytic approximations derived throughout this article.

Ultimately we are interested in the macroscopic properties of a soliton, the charge $Q$ and the energy $E$, given by the integrals
\begin{align}
    Q &= 8\pi \phi_0^2\, \omega \lambda^{3} \, \int_0^\infty \dd \rho\, \rho^2 f^2\,=\frac{4 \pi  \kappa }{d^2} \, \int_0^\infty \dd \rho\, \rho^2 f^2\,, \label{eq:Charge}\\
    E &= \int \dd^3 x \left[\phi_0^2 (f')^2 + \frac{\chi_0^2 (h')^2}{2}  + \omega^2\phi_0^2f^2 + U[\phi,\chi]\right]\nonumber\\
    &= \omega Q +\frac{8\pi}{3m_\phi} \int \dd \rho\, \rho^2 \left[\phi_0^2 (f')^2 +  \frac{\chi_0^2 (h')^2}{2}\right]\\
    &= \kappa m_\phi Q +\frac{4 \pi  m_\phi }{3 d^2} \int \dd \rho\, \rho^2 \left[(f')^2 + (h')^2\right],  \label{eq:Energy}
\end{align}
where in the last line we used the virial theorem~\cite{Derrick:1964ww,Friedberg:1976me}.

\begin{figure}[tb]
	\includegraphics[width=0.49\textwidth]{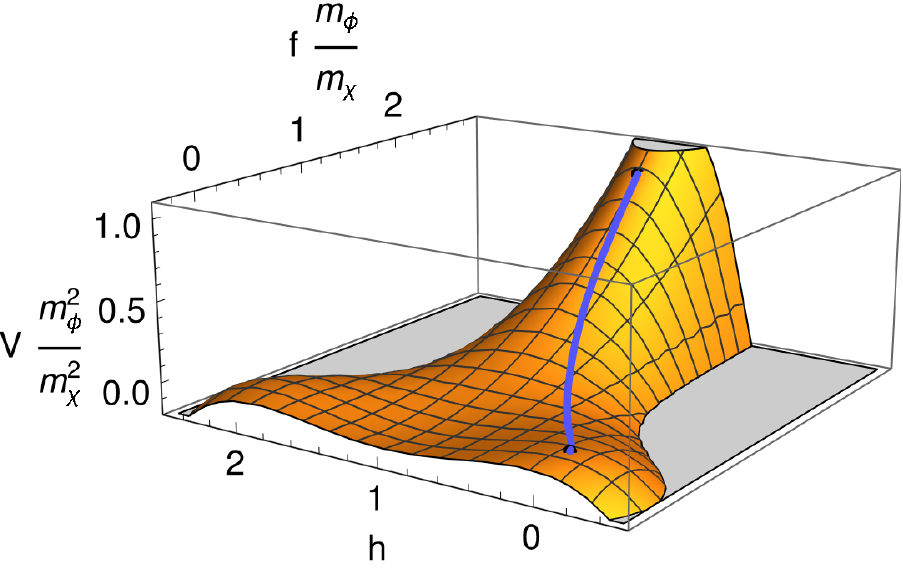}  
	\caption{ The effective potential $V$ for $\kappa=0.6$. Black dots are the start and endpoint, and the blue line the physical path (for $m_\chi=m_\phi$).
\label{fig:potential}
}
\end{figure}

The rescaled equations of motion~\eqref{eq:EOM} allow us to talk in the language of a mechanical analogy. If we treat $\rho$ as time and  $f$ and $h$ as spatial coordinates, the aforementioned system of equations 
describe the motion of a particle in the two-dimensional potential $V$, illustrated in Fig.~\ref{fig:potential}.
 In this mechanics analogy, the $2f'/\rho$ and $2h'/\rho$ terms in the differential equations correspond to time-dependent friction.
The particle starts at the point $(f(0),h(0))$ with vanishing velocity, rolls down the hill and eventually comes to rest at the local maximum $(f(\infty),h(\infty)) = (0,0)$ with vanishing potential energy. The initial potential energy is lost due to friction. From the expression for the potential $V(f,h)$ (Eq.~\eqref{eq:PotentialV}) we can see that $\kappa^2-1<0$ to ensure that $V(0,0)$ is a local maximum, thus limiting $|\kappa|$ to be less than 1. We can also restrict ourselves to positive $\kappa$ without loss of generality. The other relevant parameter, $\eta$, can take on any non-negative value, although the resulting solitons are not necessarily stable.

\section{Thin-wall limit}
\label{sec:Thin-wall}

We stress that the FLS model does not have a thin-wall limit in the sense of Coleman~\cite{Coleman:1985ki} because $V$ does not exhibit a global maximum with $V>0$ where the particle could sit and wait until the friction has died off. (This is the reason we refrain from calling the FLS solitons $Q$-balls, despite their similar properties.)
 However, there is a \emph{partial} thin-wall limit, seeing as there is a maximum at $h=1$ in the $h$ direction (Fig.~\ref{fig:potential}). The particle will still roll in the $f$ direction, but close to $h=1$ if $\kappa\ll 1$. This means that $h$ will approximately look like a step function for small $\kappa$ unless $\eta$ is also very small.  The detailed investigation of $\eta\ll1$ will come in subsection \ref{subSec:Light} but for now, we will consider  $\kappa\ll1$ and $\eta$ not too small. 
 
 For constant $h$, equation \eqref{eq:EOMExplicit_h} becomes a simple algebraic equation that has three solutions
\begin{align}
\begin{split}
    &-f^2 h+f^2-\eta h + \frac{3\eta}{2}h^2 - \frac{\eta}{2}h^3=0\,,\\
    &\Rightarrow h\in\{1,\,1-\sqrt{1-2 f^2 /\eta},\,1+\sqrt{1-2 f^2/\eta}\}\,.
    \label{eq:HMaxima}
\end{split}
\end{align}
Out of these solutions, only $h=1$ is constant with respect to $f$ so that is the value of $h$ inside the soliton. Using this step-function ansatz for $h$, we find the $f$ profile by solving  eq.~\eqref{eq:EOMExplicit_f} inside ($h=1$) and outside ($h=0$) of the $Q$-ball:
\begin{align}
\begin{split}
    &h=1 \quad \rightarrow \quad f''+\frac{2 }{\rho } f' +f \kappa ^2=0\,,\\
    &h=0 \quad \rightarrow \quad f''+\frac{2 }{\rho } f' +f \kappa ^2-f=0\,.
    \label{eq:EOMForHStep}
\end{split}
\end{align}
So far we have not determined for what value of $\rho$ does $h$ become zero. It is  natural to call that the radius of the $Q$-ball with respect to $h$ field: $R_h$. Numerically, we will use the radius definition $h''(R_h)=0$. Both inside and outside equations for $f$ are of second order so their solutions will have two free parameters. These four free parameters together with $R_h$ constitute five constants that we need to determine  for the $f$ profile. We can eliminate four of them by implementing the following boundary conditions: $f'(0)=0$, $f(\infty)=0$, $f'(R_h)[\text{inside}]=f'(R_h)[\text{outside}]$ and $f(R_h)[\text{inside}]=f(R_h)[\text{outside}]$, which gives
\begin{align}
    f = \begin{cases}
        f(0)\frac{\sin (\kappa  \rho )}{\kappa  \rho }\,, & \rho < R_h\,,\\
        f(0)\frac{1}{\rho } e^{\sqrt{1-\kappa ^2} (R_h-\rho )}\,, &\rho\geq R_h\,.
    \end{cases}
    \label{eq:FProfile}
\end{align}
where the radius of the $h$ field configuration is predicted to be
\begin{align}
    R_h= \frac{\pi -\arcsin{\kappa}}{\kappa} = \frac{\pi}{\kappa} - 1 -\mathcal{O}(\kappa^2)\,,
    \label{eq:RH}
\end{align}
which is indeed large for small $\kappa$ as expected for a thin-wall limit. The leading term in small $\kappa$ agrees with the FLS result~\cite{Friedberg:1976me}.

The two previous equations are highly reminiscent of Q-balls in flat potentials~\cite{Dvali:1997qv,Shoemaker:2009jy}. This is to be expected since the FLS potential is indeed flat in the $\phi$ direction when $\chi =\chi_0= m_\phi/d$ in Eq.~\eqref{eq:PotentialU} [or $\chi=0$ in the notation of Eq.~\eqref{eq:PotentialUFLS}]. The FLS model is hence a simple renormalizable realization of such flat-potential solitons.

The overall constant $f(0)$ is impossible to determine from here as both of the equations \eqref{eq:EOMForHStep} are linear in $f$, to be further discussed below. 
We define the radius of the $f$-field configuration by $f''(R_f)=0$, which turns out to be smaller than $R_h$ in the small-$\kappa$ regime. $f''(R_f)=0$ for $\rho < R_h$ takes the following form:
\begin{align}
    -\kappa  R_f \sin (\kappa R_f)+\frac{2 \sin (\kappa  R_f )}{\kappa  R_f }-2 \cos (\kappa   R_f)=0\,.
\end{align}
The above equation only depends on the product of $\kappa$ and $R_f$.  Because of the periodic nature of the trigonometric functions, we get infinitely many solutions $\kappa R_f\simeq2.08,\,5.94,\,9.21\dots$. The smallest positive solution yields the following radius prediction:
\begin{align}
    R_f\simeq \frac{2.08}{\kappa}\,,
    \label{eq:RF}
\end{align}
again large in the small-$\kappa$ regime and very close to $R_h$.

\begin{figure}[tb]
	\includegraphics[width=0.49\textwidth]{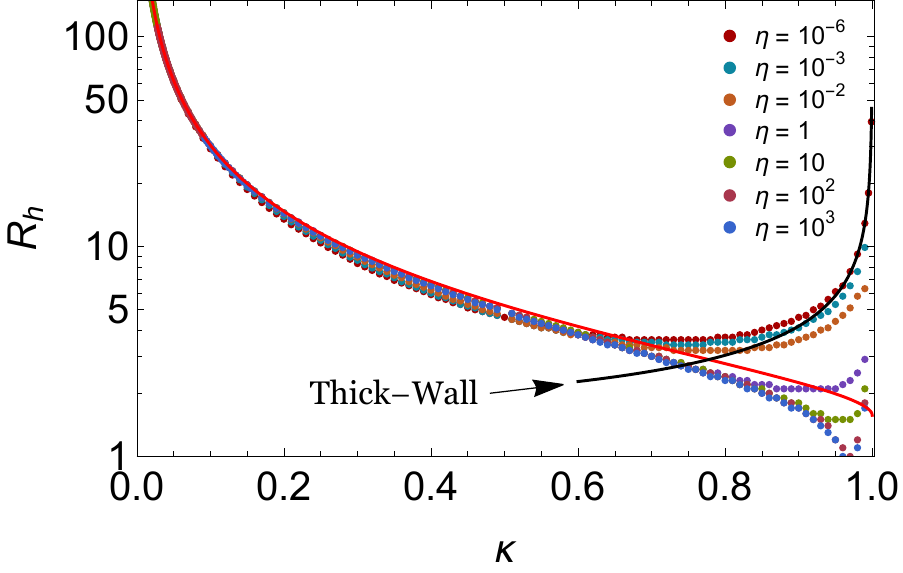}
	\caption{Radius $R_h$ as a function of $\kappa$ for several $\eta= m_\chi^2/m_\phi^2$. Dots represent numerical results while the solid red line corresponds to our thin-wall prediction from Eq.~\eqref{eq:RH}. The solid black line shows the thick-wall small-$\eta$ prediction from Eq.~\eqref{eq:thick_radius}.
 }

	\label{fig:PlotRadii}
\end{figure}

Note that even though our theory depends on two parameters, both of our radius predictions are independent of $\eta$ and only depend on $\kappa$. We verify this by comparing $R_h$ to the numerical results in Fig.~\ref{fig:PlotRadii}. One can also notice that while the theoretical and numerical results are in agreement for every $\eta$ in the thin-wall regime $(\kappa\ll1)$, this agreement only holds for large $\eta$ values as $\kappa$ increases. This is due to our step-function approximation for the $h$ field. As we will see later in section \ref{subSec:heavy}, this approximation holds best for large $\eta$ regime, as can be seen already in Fig.~\ref{fig:PlotProfiles}.

With the $f$ profile from eq.~\eqref{eq:FProfile} at our disposal it is easy to calculate two out of the three integrals that are necessary for calculating soliton energy and charge. We can simply plug in and integrate our results from eq.~\eqref{eq:FProfile}:
\begin{align}
    &\int\limits_0^\infty \dd \rho\, \rho^2 f^2 =f(0)^2\frac{ \left(2 \cos ^{-1}(\kappa )+\pi +\frac{2 \kappa }{\sqrt{1-\kappa ^2}}\right)}{4 \kappa ^3}\,,\nonumber\\
    &\int\limits_0^\infty \dd \rho\, \rho^2 f'^2 =f(0)^2\frac{ \left(2 \cos ^{-1}(\kappa )+\pi \right)}{4 \kappa }\,.
    \label{eq:Integrals}
\end{align}
At small $\kappa$, the two integrals differ only by a factor $1/\kappa^2$.

To obtain an estimate for the last unknown parameter $f(0)$, we return to the mechanical analogy described at the beginning of this section. A stationary particle starts motion at $f=f(0)\,\&\,h\simeq 1$ and stops at $f=0=h$. The change in potential (and total mechanical) energy is
\begin{align}
    V(0,0)-V(f(0),1)=-\frac{1}{8} \left[4 f(0)^2 \kappa ^2-\frac{m_\chi^2}{m_\phi^2} \right].
    \label{eq:DeltaV}
\end{align}
This energy is lost due to time-dependent friction forces in $f$ and $h$ directions. The total work done by them is
\begin{align}
    W=-\int \dd\rho \, \frac{2}{\rho} \left[ (f')^2+(h')^2\right].
    \label{eq:Work}
\end{align}
The integral over $(f')^2$ can be performed with our ansatz, but we need to go beyond the step-function approximation for $h$ to determine the integral over $(h')^2$. After that, $W=\Delta V$ reduces to a simple algebraic equation with respect to $f(0)$ and can be solved easily. As we will see later, $h$ exhibits qualitatively different behavior in large (heavy $\chi$) and small (light $\chi$) $\eta$ cases. Because of that, we will have to derive two separate predictions of $f(0)$ to characterize the entire range of $\eta$.
 
\subsection{Heavy \texorpdfstring{$\chi$}{chi}}
\label{subSec:heavy}

\begin{figure}[tb]
	\includegraphics[width=0.49\textwidth]{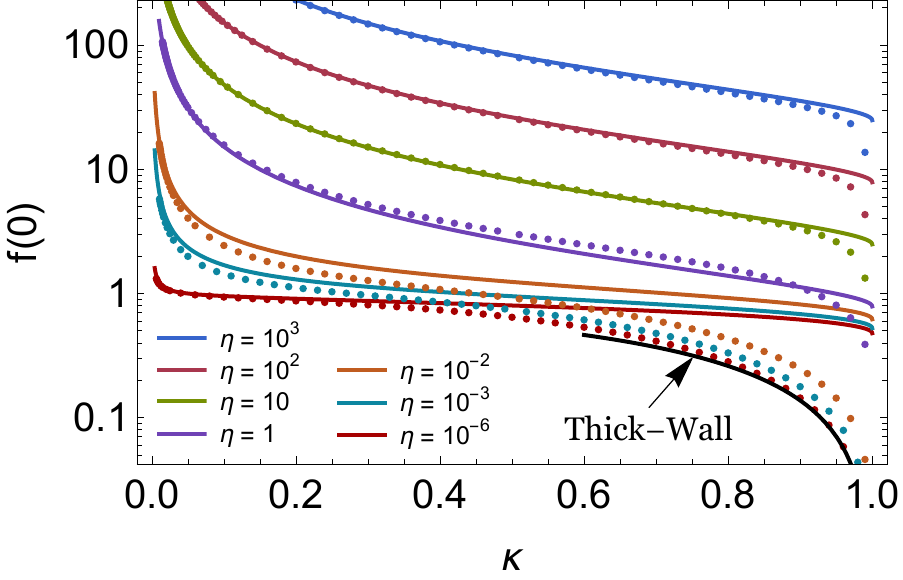} 
	\caption{ $f(0)$ as a function of $\kappa$ for several $\eta= m_\chi^2/m_\phi^2$. Dots represent numerical results while solid lines correspond to our theoretical predictions, using Eq.~\eqref{eq:F0Heavy} for $\eta \geq 1$ and Eq.~\eqref{eq:F0Light} for $\eta < 1$.
}
	\label{fig:PlotF0}
\end{figure}

According to the numerical results, for large $\eta$ values the particle sits at the maximum with respect to $h$ and rolls exclusively in the $f$ direction for the majority of the motion. The particle only starts moving along $h$ at the very end of the trajectory when due to the $\frac{2}{\rho}$ prefactor the friction has already diminished to an inconsequential level. To illustrate that the energy loss due to $h$ friction can be ignored, let us try to approximate it from above. From the equation~\eqref{eq:HMaxima} we can see that for $f>\sqrt{\frac{\eta}{2}}$, the potential $V$ has only one extremum and that is its global maximum at $h=1$. As $f$ drops below $\sqrt{\frac{\eta}{2}}$ it splits this extremum into three: one minimum (at $h=1$) and two maxima (at $h=0$ and $2$). As we already mentioned, we are going to ignore solutions that roll toward the $h>1$ direction. For the particle to end up at $h=0$, it needs to stay near that maximum throughout the motion. Otherwise, depending on the direction of the displacement, it will return to $h=1$ minima and fall in the direction of $h\approx1$ \& $f\rightarrow -\infty$  or start accelerating uncontrollably and roll somewhere in the $h\rightarrow -\infty$ direction. Either way, we are not going to recover a localized solution. This leaves us with the following estimate, where $h$ tracks the maximum:
\begin{align}
    h = \begin{cases}
        1, & \rho < R_m\,,\\
        1-\sqrt{1-2 f^2 /\eta}, &\rho\geq R_m\,.
    \end{cases}
    \label{eq:HProfileHeavy}
\end{align}
Here, $R_m$ is the value of $\rho$ for which $1-2 f(R_m)^2 /\eta=0$ and the particle starts moving in the $h$ direction. 
The above shape for $h$ is remarkably good for large $\eta$.
As we are considering the large $\eta$ regime we can keep only the leading term in the Eq.~\eqref{eq:HProfileHeavy}, $h = 1-\sqrt{1-2 f^2 /\eta} \approx \frac{f^2}{\eta}$. We can now proceed to estimate the  $h$ integral contribution to $W$:
\begin{align}
\begin{split}
    \int\limits_{0}^\infty \dd \rho\, \frac{2}{\rho} (h')^2 \approx \int\limits_{R_m}^\infty \dd \rho\, \frac{2}{\rho} \left( \frac{2 f' f}{\eta} \right)^2 \leq \frac{2}{\eta} \int\limits_{R_m}^\infty \dd \rho\, \frac{2}{\rho} (f')^2.
    \label{eq:InvIntHPHeavy}  
\end{split}
\end{align}
Here, in the last step, we replaced $f(\rho)$ with $f(R_m)$. 
Since we expect  $R_m \approx R_h$, it has a $1/\kappa$ dependence. As we anticipated above, the integral is suppressed in the thin-wall limit as it starts at some large value of $\rho$ and has a $2/\rho$ dependence. On top of that, now we can see that it is also suppressed in the heavy-$\chi$ limit by the factor of $2/\eta$. All of this allows us to neglect the $h$ contribution in the work integral $W$. With the $f$ profile (Eq.~\eqref{eq:FProfile}) at our disposal we then find 
\begin{align}
\begin{split}
    &W=-\frac{1}{2} f(0)^2 \kappa ^2 \left[1-\frac{4}{\left(2 \cos ^{-1}(\kappa )+\pi \right)^2}\right].
\end{split}
\end{align}
Setting this equal to the potential difference we get
\begin{align}
   f(0)=\sqrt{\eta}\;\frac{\pi -\sin ^{-1}(\kappa )}{2 \kappa }
   \label{eq:F0Heavy}
\end{align}
for large $\eta$.
The $\kappa$ dependence together with Eq.~\eqref{eq:Integrals} satisfies $\dd E/\dd \omega = \omega \dd Q/\dd \omega$, which should be valid for the true solution~\cite{Friedberg:1976me,Lee:1991ax}.
This prediction for $f(0)$ is compared to the numerical data in Fig.~\ref{fig:PlotF0}. As expected, Eq.~\eqref{eq:F0Heavy} predicts numerical results best for $\kappa\ll 1\: \& \: \eta\gg 1$ region. Unexpectedly, for $\eta\gtrsim 1$ this agreement holds even outside the thin-wall limit all the way up to $\kappa \approx 0.85$. 

\begin{figure}[tb]
	\includegraphics[width=0.49\textwidth]{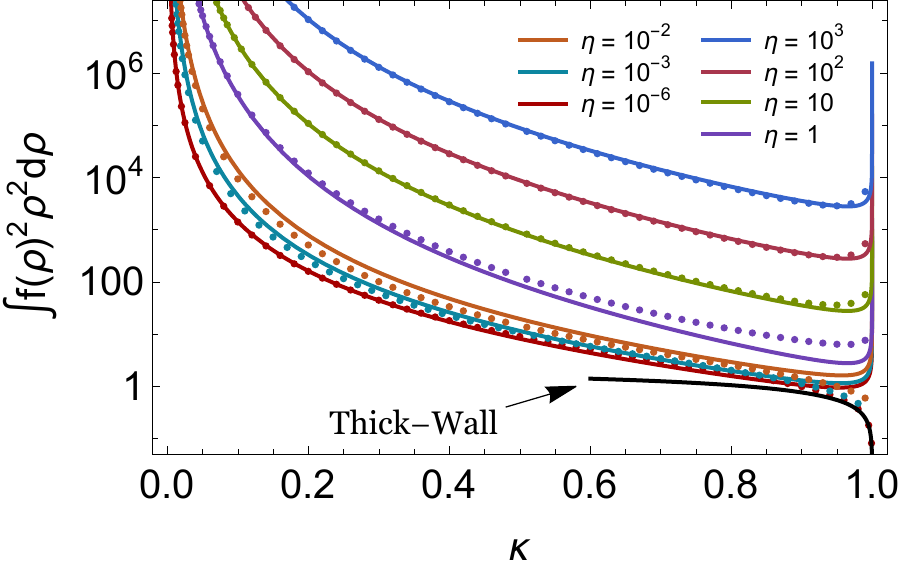}  \\
	\includegraphics[width=0.49\textwidth]{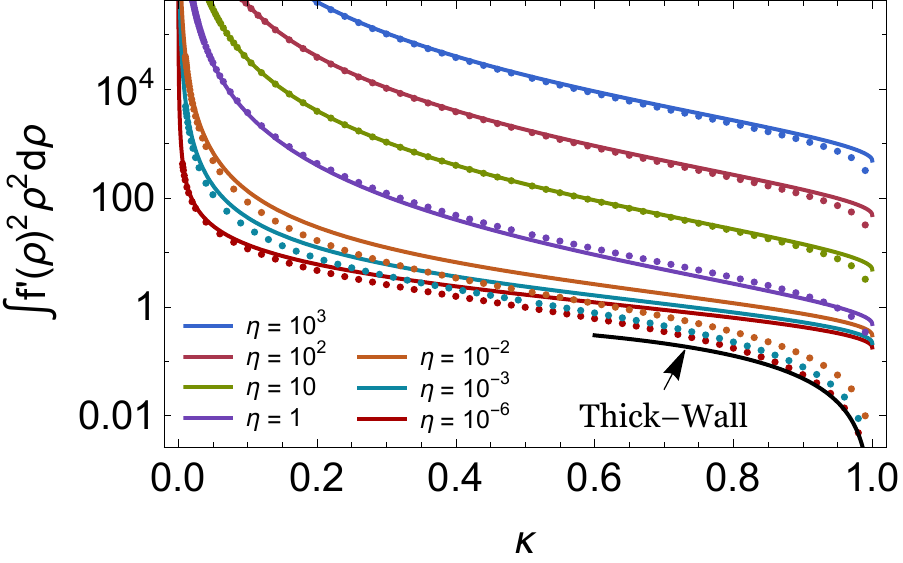}\\
	\includegraphics[width=0.49\textwidth]{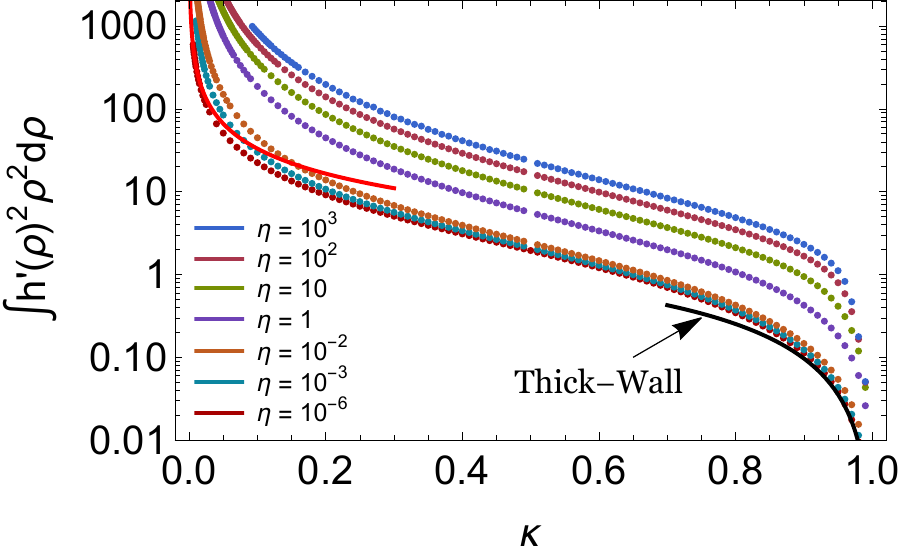}
	\caption{ $\int [f(\rho)]^2\rho^2\dd\rho $ (top), $\int [f'(\rho)]^2\rho^2\dd\rho $ (middle) and $\int [h'(\rho)]^2\rho^2\dd\rho $ (bottom) as functions of $\kappa$ for several $\eta= m_\chi^2/m_\phi^2$. Dots correspond to the numerical values, while colorful solid lines show our thin-wall approximations from Eq.~\eqref{eq:Integrals}. The black lines represent the $\eta\rightarrow 0$ limit of the thick-wall regime for all three figures and the red line in the third plot illustrates the thin-wall regime in the same $\eta\rightarrow 0$ limit. For $f(0)$, we used Eq.~\eqref{eq:F0Heavy} for $\eta \geq 1$ and Eq.~\eqref{eq:F0Light} for $\eta < 1$.
 }
 \label{fig:PlotIntegrals}
\end{figure}

Equation~\eqref{eq:F0Heavy} together with equations~\eqref{eq:Integrals} allow us to provide analytical expressions for the aforementioned integrals, 
\begin{align}
    \int_0^\infty \dd \rho\, \rho^2 f^2 &= \frac{\pi^3\eta}{8\kappa^5} +\mathcal{O}(\kappa^{-4})\,,\\    \int_0^\infty \dd \rho\, \rho^2 f'^2 &= \frac{\pi^3\eta}{8\kappa^3} +\mathcal{O}(\kappa^{-2})\,,
\end{align}
at leading order in small $\kappa$.
The full theoretical results are represented by the solid lines in Fig.~\ref{fig:PlotIntegrals}. Here we notice a pattern similar to the one observed in Fig.~\ref{fig:PlotF0}, there is an excellent agreement in the thin-wall-heavy-$\chi$ regime, which still holds well beyond the thin-wall region all the way up to $\kappa \approx 0.85$ for $\eta\gtrsim 1$. It is also worth mentioning that our estimate of the non-primed integral is more accurate compared to the primed one. That is due to the fact that the latter is more sensitive to the imperfections of the $f$-profile prediction (Eq.~\eqref{eq:FProfile}).
Note that we could use Eq.~\eqref{eq:HProfileHeavy} to obtain a theoretical prediction for the $(h')^2$ integral, but since this one is negligible compared to the $(f')^2$ integral for large $\eta$ we will not bother.

\subsection{Light \texorpdfstring{$\chi$}{chi}}
\label{subSec:Light}

For $m_\chi \ll m_\phi$, the friction in $h$ direction is no longer negligible and we need a different approximation. We can neglect all the terms that contain second or higher order of $h$ in Eq.~\eqref{eq:EOMExplicit_h} and only keep the linear contribution. Even the term $f^2 (1-h)$ is no longer significant, as it is suppressed by $f^2$ outside the $Q$-ball and by $h\simeq 1$ inside. That leaves us with
\begin{align}
    h''+\frac{2}{\rho}h'-\eta h = 0\,.
    \label{HEOMLinear}
\end{align}
It is then easy to find that $h=1$ inside and $h\propto e^{-\sqrt{\eta}\rho}/\rho$ outside. We yet have to decide for what value of $\rho$ should we match these solutions. From our previous results (Eq.~\eqref{eq:RH} and Eq.~\eqref{eq:RF}) it is clear that the transition region is somewhere around $\rho \propto 1/\kappa$. For now, let's keep the $\mathcal{O}(1)$ proportionality constant arbitrary and call it $C_\text{Tran}$, where ``Tran'' stands for transition. Then to make the two solutions coincide at $\rho = C_\text{Tran}/\kappa$ we should choose the coefficient of the outside equation accordingly, 
\begin{align}
    h(\rho)_\text{outside} = \frac{\frac{C_\text{Tran}}{\kappa} \, e^{\sqrt{\eta} \, \frac{C_\text{Tran}}{\kappa} }}{\rho \, e^{\sqrt{\eta}\rho}}\,.
    \label{eq:HProfileLight}
\end{align}
Now we can calculate the $h$ contribution to $W$ and find $f(0)$ for the Light-$\chi$ case, just like we did for the heavy one. Combining equations \eqref{eq:DeltaV}, \eqref{eq:Work}, \eqref{eq:FProfile}, and \eqref{eq:HProfileLight} yields the following expression:
\begin{align}
    f(0)=\frac{\sqrt{C_{\text{Tran}}^2 \eta + 8\, C_{\text{Tran}} \sqrt{\eta }\, \kappa  +4 \kappa ^2}\left(\pi  + 2 \cos ^{-1}\kappa \right)}{4\, C_{\text{Tran}}\, \kappa} \,.
    \label{eq:F0Light}
\end{align}
We fix the arbitrary parameter $C_{\text{Tran}}=3.3$ by fitting the prediction of $f(0)$ to numerical results in $\eta = 10^{-6}$ case.
We can also estimate another $h$-integral that is needed to calculate the energy (Eq.~\eqref{eq:Energy}). With this profile, the integral simplifies as follows,

\begin{align}
\begin{split}
\lim_{\eta\to 0} \int\limits_0^\infty \dd \rho\, \rho^2 h'^2 &= \lim_{\eta\to 0} \int\limits_{\frac{C_\text{Tran}}{\kappa}}^\infty \dd \rho\, \rho^2 h'^2 \\
&= \lim_{\eta\to 0} \frac{C_\text{Tran} \left(2 \kappa + C_\text{Tran}\sqrt{\eta }\right)}{2 \kappa ^2}\\
&=\frac{C_\text{Tran}}{\kappa} \,.
\label{eq:IntHPLight}
\end{split}
\end{align}
This result is represented by the solid red line in the bottom plot of Fig.~\ref{fig:PlotIntegrals}. It follows a general shape of the numerical plots and exhibits the same behavior as $\eta = 10^{-6}$ case in $\kappa \to 0$ limit.

\section{Thick-wall limit}
\label{sec:Thick-wall}

In the limit $\kappa\to 1$, the fields have suppressed amplitudes, especially for small $\eta$ (see Fig.~\ref{fig:PlotProfiles}). This allows us to neglect all cubic terms in the differential equations \eqref{eq:EOMExplicit_f} and \eqref{eq:EOMExplicit_h}. We restrict ourselves to the small $\eta$ regime because only there do we have stable $Q$-balls with $\kappa \sim 1$. All of this is analogous to Kusenko's thick-wall $Q$-balls~\cite{Kusenko:1997ad}. Neglecting $\eta$, we can make the ansatz
\begin{align}
\begin{split}
f(\rho) &= (1-\kappa^2) \,\, f_\text{Th}\left( \sqrt{1-\kappa^2} \rho\right) ,\\
h(\rho) &= (1-\kappa^2) \,\, h_\text{Th}\left( \sqrt{1-\kappa^2} \rho\right) ,
\label{eq:ProfilesThick}
\end{split}
\end{align}
where $f_\text{Th}$ and $h_\text{Th}$ are yet again rescaled fields. By plugging them into the differential equations we recover a  set of parameterless differential equations:
\begin{align}
&f''_{\text{Th}}(P)+\frac{2 f'_{\text{Th}}(P)}{P}+2 f_{\text{Th}}(P) h_{\text{Th}}(P)-f_{\text{Th}}(P)=0 \,,\nonumber\\
&h''_{\text{Th}}(P)+\frac{2 h'_{\text{Th}}(P)}{P}+f^2_{\text{Th}}(P)=0 \,,
\label{eq:EOMThick}
\end{align}
where $P\equiv\sqrt{1-\kappa^2}\rho$ and the derivatives are taken with respect to $P$. We only need to solve these equations once numerically. After, we can use those solutions to generate any profiles in the thick-wall-light-$\chi$ approximation using Eq.~\eqref{eq:ProfilesThick}. One example ($\kappa = 0.95$ \& $\eta = 10^{-6}$) of these profiles is shown in Fig.~\ref{fig:PlotProfiles}.

The soliton radii in the thick-wall regime are
\begin{align}
    R_h \simeq \frac{1.82}{\sqrt{1-\kappa^2}}\,, &&
    R_f \simeq \frac{1.68}{\sqrt{1-\kappa^2}}\,, 
    \label{eq:thick_radius}
\end{align}
which diverge for $\kappa\to 1$.
This thick-wall limit also allows us to calculate the necessary integrals for $E$ and $Q$, they take the simple form
\begin{align}
\int_{}^{} [f'(\rho)]^2\rho^2\dd\rho
&= \tfrac12 \int_{}^{} [h'(\rho)]^2\rho^2\dd\rho\\
&= \tfrac13 (1-\kappa^2)\int_{}^{} [f(\rho)]^2\rho^2\dd\rho\\
&= (1-\kappa^2)^{\frac{3}{2}} \times 0.584\,.
\label{eq:IntegralsThick}
\end{align}
The soliton charge $Q$ vanishes for $\kappa\to 1$, as do the field amplitudes, while the radius diverges, indicating the transition to the vacuum solution for $\kappa\to 1$.
However, our classical-field description
of these solitons eventually breaks down at small
$Q$ and needs to be replaced by a quantum-mechanical picture in which $Q$ takes on integer values only~\cite{Graham:2001hr}. Notice that while $Q$ can indeed be small near $\kappa \sim 1$, its actual value depends on the otherwise-irrelevant potential parameter $d$, see Eq.~\eqref{eq:Charge}.
The above results are represented with solid black lines in figures \ref{fig:PlotIntegrals} and \ref{fig:PlotF0}. As expected, in the $\kappa \sim 1$ region numerical values converge on our theoretical thick-wall-light-$\chi$ line as $\eta$ becomes smaller and smaller. Additionally, there is a reasonable agreement with $\eta \leq 1$ cases as $\kappa$ approaches one.

It is worth mentioning that
for finite $\eta$, the integral over $f^2$ and hence $Q$ eventually start to \emph{increase} as $\kappa\to 1$, although that is difficult to see in Fig.~\ref{fig:PlotIntegrals} for the small-$\eta$ cases.
 FLS~\cite{Friedberg:1976me} have derived a different thick-wall prediction, one that relies on $(1-\kappa^2) \ll \eta$ and properly describes this behavior. We omit a discussion of this approximation here since it only applies to the parameter space where the solitons are unstable, as emphasized in the next section.

\section{Stability}
\label{sec:stability}

The solitons are stable if they have the lowest energy among all field configurations with the same charge $Q$. In particular, we need $E/(m_\phi Q) < 1$ in order to evade soliton decay into $Q$ individual scalars. With our definitions, $E/(m_\phi Q)$ is a function of $\kappa$ and $\eta$, 
\begin{align}
\frac{E}{m_{\phi}Q}=\kappa+\frac{1}{3\kappa}\frac{\int\limits_0^\infty \dd \rho\, \rho^2 f'^2+\int\limits_0^\infty \dd \rho\, \rho^2 h'^2}{\int\limits_0^\infty \dd \rho\, \rho^2 f^2 }\,,
\label{eq:Emqgeneral}
\end{align}
so we can determine the stable region in the  $\kappa$--$\eta$ plane.
Let us consider this problem in the three different limits studied above. Using equations \eqref{eq:Integrals} and \eqref{eq:IntHPLight} in the thin-wall small-$\eta$ regime the ratio reduces to
\begin{align}
\lim_{\eta\to 0} \, \frac{E}{m_{\phi}Q}=\frac{4}{3} \kappa  \left(1+\frac{ C^3_\text{Tran}}{2\pi ^3}\right)+\mathcal{O}\left(\kappa ^2\right) ,
\label{eq:EmqThinLight}
\end{align}
which is indeed smaller than 1 for sufficiently small $\kappa$.
Hence $E<m_{\phi}Q$ and the soliton is stable in the thin-wall-light-$\chi$ limit. On the other hand, in the thin-wall-\emph{heavy}-$\chi$ regime, the $h$-integral is negligible \footnote{It is negligible with the same logic as the $h$-integral in equation Eq.~\eqref{eq:InvIntHPHeavy}. In the current case we will not have the same thin-wall ($\kappa \ll 1$) suppression but the $1/\eta$ dependence will still be there.} and the $f(0)$ dependence completely drops out, leaving us with only the $\kappa$ dependence. Expanding the ratio near $\kappa \sim 0$, we get
\begin{align}
\frac{E}{m_{\phi}Q}=\frac{4 }{3} \kappa-\frac{\kappa ^2}{3 \pi }+\mathcal{O}\left(\kappa ^3\right) ,
\label{eq:EmqThinHeavy}
\end{align}
again smaller than 1 for small $\kappa$.
Thus, the thin-wall-heavy-$\chi$ regime also yields stable solitons. Finally, if we use our thick-wall-light-$\chi$ equations (Eq.~\eqref{eq:IntegralsThick}) and expand in the vicinity of $\kappa \sim 1$, we get
\begin{align}
\frac{E}{m_{\phi}Q}\simeq1-\frac{1-\kappa}{3} +\mathcal{O}((1-\kappa)^2)\,,
\end{align}
rendering these $\kappa\sim 1$ solitons stable for small $\eta$, although with a smaller binding energy than in the thin-wall cases.

\begin{figure}[tb]
	\includegraphics[width=0.49\textwidth]{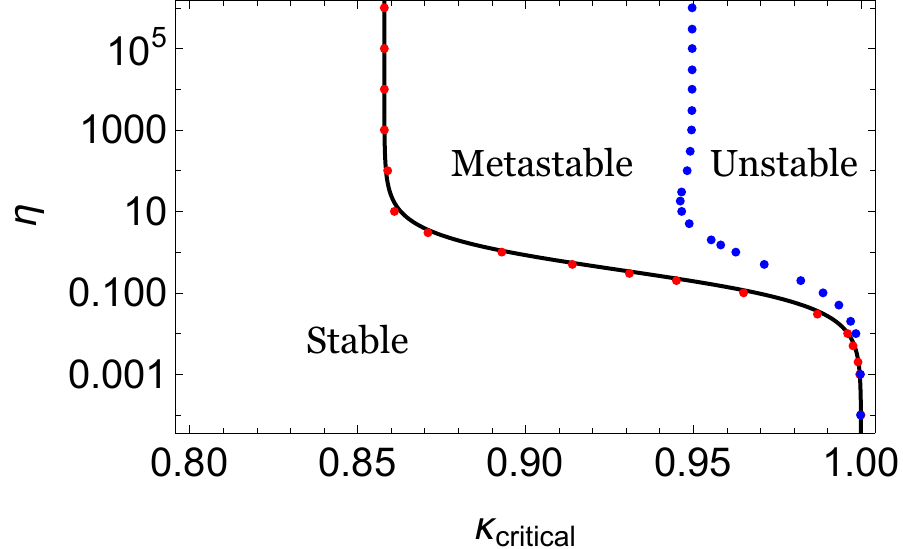}
	\caption{ Stability diagram in the $\eta$--$\kappa$ plane.
	Red and blue dots represent numerical results and the black line is the empirical relation from Eq.~\eqref{eq:EtaVsKappaCriticalEmp}. $E< m_\phi Q$ in the ``Stable'' region, see text for details.
}
	\label{fig:PlotEtaVsKappaCritical}
\end{figure}

All of these results are in qualitative agreement with the numerical data illustrated in Fig.~\ref{fig:PlotEtaVsKappaCritical}. Here we have defined $\kappa_\text{critical}$ so that every $\kappa > \kappa_\text{critical}$ yields an unstable soliton solution, i.e.~one with $E>m_{\phi}Q$. The red dots are generated numerically,
the black line shows the empirically successful relation
\begin{align}
\eta = \frac{1 - 0.858}{\kappa_\text{critical} - 0.858}\frac{1-\kappa_\text{critical}}{0.4} \,,
\label{eq:EtaVsKappaCriticalEmp}
\end{align}
which exhibits the correct behavior in both the $\kappa \to 1$ and $\kappa \to 0.858$ limits. This empirical equation 
 is in excellent agreement with the numerical results.

From Fig.~\ref{fig:PlotEtaVsKappaCritical} and Fig.~\ref{fig:PlotRadii} we can now also conclude that \emph{stable} FLS solitons have radii $R_h >1$, or 
\begin{align}
R_\text{soliton} > 1/m_\phi
\end{align}
in dimensionfull units, in perfect agreement with the bound-state conjecture of Ref.~\cite{Freivogel:2019mtr}.

Note that while $E< m_\phi Q$ indeed guarantees soliton stability, solitons with $E> m_\phi Q$ can still be stable at the \emph{classical} level: only for $\dd Q/\dd \omega >0$ can solitons decay/fission into smaller solitons with lower energy, which clearly renders them unstable~\cite{Friedberg:1976me,Tsumagari:2008bv} (see Fig.~\ref{fig:PlotEtaVsKappaCritical}). 
In between, there lies a region with $E> m_\phi Q$ and  $\dd Q/\dd \omega <0$ in which the solitons can lower their energy by dispersing into individual scalars through quantum effects; since this is likely parametrically slower than classical decay, these have been denoted as metastable in Ref.~\cite{Friedberg:1976me}.

\section{Discussion and comparison to \texorpdfstring{$Q$-balls}{Q-balls}}
\label{sec:discussion}

We found the following general properties of FLS solitons: i) The internal structure only depends on two parameters, $\kappa \equiv \omega / m_\phi$ and $\eta\equiv m_\chi^2/m_\phi^2$, defined just before and after the Eq.~\eqref{eq:EOM}. ii) For large solitons, radii with respect to each of the fields grow with $1/\kappa$ up to an $\mathcal{O}(1)$ prefactor that only weakly depends on the mass ratio. iii) The stable region is also determined by the two parameters $\eta$ and $\kappa$. We explained theoretically why we expect stable solutions in the following three limits: $\kappa \to 0$ \& $\eta \to 0$, $\kappa \to 0$ \& $\eta \to \infty$, and $\kappa \to 1$ \& $\eta \to 0$ and confirmed our results numerically.

For small $\kappa$ and large $\eta$, we can eliminate $\kappa$ from our analytical approximations and obtain the following relationships between the macroscopic soliton properties: 
\begin{align}
    Q &= \frac{\eta }{2 d^2} \, R_h^4 \left(1+\frac{2}{R_h}+\frac{1}{R_h^2} + \mathcal{O}(R_h^{-3}) \right),\label{eq:Q_large_eta}\\
    E &= \frac{2^{7/4} \pi  \eta^{1/4}  m_\phi}{3 \sqrt{d}} \, Q^{3/4} \left(1- \frac{3   \eta^{1/4} }{ 2^{9/4} \sqrt{d} Q^{1/4}} + \mathcal{O}(Q^{-\tfrac12}) \right) ,\nonumber
\end{align}
where the solitons become arbitrary large. The leading-order terms agree with the expressions in Ref.~\cite{Friedberg:1976me} that was obtained from a variational argument.
The scaling of $Q$ with the radius is markedly different from that of Coleman's $Q$-balls, for which both $Q$ and $E$ simply scale with the soliton volume. This difference can be traced back to the absence of a full thin-wall limit in the FLS model, i.e.~the fact that the $f$ profile is never constant.
From the above $E\propto Q^{3/4}$ expression --  which matches the expectation for Q-balls in flat potentials~\cite{Dvali:1997qv,Shoemaker:2009jy} -- it is obvious that the solitons are stable against dispersion once the charge exceeds a critical value~\cite{Friedberg:1976me},
\begin{align}
    Q_\text{critical} \simeq \frac{128\pi^4\eta}{81 d^2} \,,
\end{align}
seeing as $Q$ free $\phi$ particles have energy $m_\phi Q$ that grows much faster with $Q$.

In the thin-wall-\emph{light}-$\chi$ regime, the leading-order terms are the same as in Eq.~\eqref{eq:Q_large_eta}.
Finally, in the thick-wall-light-$\chi$ limit, we have
\begin{align}
    Q &= \frac{12\pi\, 0.584}{d^2}\, \frac{1.82}{R_h}\left[1- \frac{1}{2}\left(\frac{1.82}{R_h}\right)^2+\mathcal{O}(R_h^{-4})\right],\\
    E &= m_\phi Q \left[1 - \frac{d^4 Q^2}{864\pi^2 \,0.584^2}+\mathcal{O}(Q^{4})\right] ,
\end{align}
where $R_h(\kappa)$, given by Eq.~\eqref{eq:thick_radius}, is large, so both $Q$ and $E$ are small.
The thick-wall behavior of the FLS solitons is qualitatively similar to that of thick-wall $Q$-balls~\cite{Kusenko:1997ad,Heeck:2022iky}.

\section{Conclusion}
\label{sec:conclusion}

In this article, we investigated one of the earliest and simplest realizations of non-topological solitons, that of Friedberg, Lee, and Sirlin~\cite{Friedberg:1976me}.
These two-field solitons do not admit a thin-wall limit in the sense of Coleman and are hence different from the more commonly studied $Q$-balls, despite many similarities.
Even without Coleman's thin-wall limit, the FLS solitons form stable, arbitrarily large objects as long as their charge exceeds a minimal value.
We have provided analytical approximations that describe these stable solitons for wide range of all free parameters and presented corresponding numerical calculations that go beyond the range of validity of analytic solutions to confirm our claims.

\vspace{2ex}

\section*{Acknowledgements}
This work was supported in part by the National Science Foundation under Grant PHY-2210428.
Numerical data files can be found as ancillary files on the arXiv page of this article~\cite{Heeck:2023idx}.

\bibliographystyle{utcaps_mod}
\bibliography{BIB}

\end{document}